\documentstyle[12pt]{article}
\input /users/lpthe/keum/feynman.tex

\newcommand{\beq}{\begin{equation}}
\newcommand{\eeq}{\end{equation}}
\newcommand{\beqa}{\begin{eqnarray}}
\newcommand{\eeqa}{\end{eqnarray}}
\newcommand{\beqan}{\begin{eqnarray*}}
\newcommand{\eeqan}{\end{eqnarray*}}
\newcommand{\no}{\nonumber}

\newcommand{\ul}{\underline}
\newcommand{\ol}{\overline}
\newcommand{\ra}{\rightarrow}
\newcommand{\lra}{\longrightarrow}
\newcommand{\Lra}{\Longrightarrow}

\newcommand{\ben}{\begin{enumerate}}
\newcommand{\een}{\end{enumerate}}
\newcommand{\bfl}{\begin{flushleft}}
\newcommand{\efl}{\end{flushleft}}
\newcommand{\ba}{\begin{array}}
\newcommand{\ea}{\end{array}}
\newcommand{\btab}{\begin{tabular}}
\newcommand{\etab}{\end{tabular}}
\newcommand{\bit}{\begin{itemize}}
\newcommand{\eit}{\end{itemize}}

\newcommand{\vs}{\vspace}
\newcommand{\hs}{\hspace}

\newcounter{muni}

\begin{document}

\vspace{.4cm}
\hbadness=10000
\pagenumbering{arabic}
\pagestyle{myheading}
\title{
FAILURE OF THE COMMONLY USED $B \ra K ( K^* )$ FORM FACTORS \\
IN EXPLAINING RECENT DATA ON $B \ra J/\Psi + K ( K^* ,$ DECAYS }
\author{$\ $\\
{\bf  M. Gourdin$^1$, A. N. Kamal$^{1,2}$ and X. Y. Pham$^1$} \vs{10mm} $\ $\\
$^1$Laboratoire de Physique Th\'eorique et Hautes Energies,\thanks{\em
Unit\'e  associ\'ee au CNRS URA 280, \rm\newline $1$ \hspace{0.2cm}
Postal address: LPTHE, Tour 16, $1^{er}$ Etage, Universit\'e Pierre et Marie
Curie,
 4 Place Jussieu,
\newline \hs{22mm}
F-75252 Paris CEDEX 05 France.
\newline
$2$ \hs{0.2cm} Permanent address : Department of Physics, University of
Alberta,
 Edmonton, Alberta T6G2JI,
\newline \hs{22mm} Canada }~ Paris, France \\
\vs{10mm}
{\small  Email: gourdin@lpthe.jussieu.fr, kamal@lpthe.jussieu.fr and
pham@lpthe.jussieu.fr } }
\date{}   
\maketitle
\vs{-11cm}
\begin{flushright}
{\bf
PAR/LPTHE/94--19 \\
May 11, 1994 \\ }
\end{flushright}

\thispagestyle{empty}
\vs{110mm}

\begin{abstract}
\noindent{
 We establish restrictions on $B \ra K ( K^* )$ form factors imposed by recent
data on $B \ra J/\Psi + K ( K^* ) $ decay rates and polarization.
We show that these constraints are not satisfied by few commonly used models.

Finally, we relate $B \ra K ( K^* )$ form factors to those in $D \ra K ( K^* )$
transitions using the heavy flavour symmetry as proposed by Isgur and Wise and
discuss the uncertainties in this procedure. We find that this method also
leads to form factors excluded by data.

\end{abstract}

\vs{35mm}
{\bf PACS index 11.30. Hv, 13. 25. Hw, 14. 40. Nd }

\newpage
\vs{5mm}
It is generally believed that the best place to study the importance of
colour-suppressed processes in \it B \rm  meson decay is to look at final
states which, at the quark level, have the form

\beq
( \ol{b}q) \hs{2mm} \Lra \hs{2mm} ( \ol{c}c) + ( \ol{s}q ), \hs{20mm} q = u, d
\label{eq1}
\eeq

The corresponding colour-suppressed diagram is shown in Figure 1.


\begin{center}
$$
\beginturtle
\shift{-150pt}{0pt}
\arrowtrue
\rotate{180}\putclabel{$\ol{b}$} \jump{-10pt}
\solidline{-70pt} \solidline{-70pt} \jump{-10pt} \putclabel{$\ol{c}$}
\jump{80pt} \rotate{135} \wavyline{40pt}{8}
\rotate{60} \solidline{40pt} \jump{10pt} \putclabel{$c$} \jump{-50pt}
\rotate{150} \solidline{-40pt} \jump{-10pt} \putclabel{$\ol{s}$}
\rotate{195} \shift{0pt}{-10pt} \putclabel{$q$}
\jump{-10pt} \solidline{-70pt}
\solidline{-70pt} \jump{-10pt} \putclabel{$q$}
\endturtle
$$

\vs{2mm}
Figure 1.
\end{center}

The relevant Cabibbo-Kobayashi-Maskawa (CKM) factor is $V_{cb}^{*}V_{cs}$ and,
of course, penguin diagrams also contribute to these decays.
However as the colourless charmonium state $\ol{c}c$ has to be excited from the
vacuum, at least three gluons are needed for the $J/\Psi$ and  $\Psi(2S)$  and
two for the $\chi_{C1}$(1P).
For this reason the penguin contribution will be neglected in this letter.

The charmonium states that have been observed experimentally [1] are $J/\Psi$,
$\Psi(2S)$ and $\chi_{C1}$(1P) along with the strange mesons $K$ and
$K^{*}(892)$ in both neutral and charged modes.
We restrict our discussion to the following processes

\beqa
B^{o}_d \hs{2mm} & \Lra & \hs{2mm} J/\Psi + K^o ( K^{*o} )  \no \\
& & \no \\
B^{+}_u \hs{2mm} & \Lra & \hs{2mm} J/\Psi + K^{+} ( K^{*+} ) \label{eq2}
\eeqa

The computation of the branching ratios for the decays in (2) involves the
lifetime $\tau_B$, the CKM factor, the leptonic decay constant $f_{J/\Psi}$ and
the hadronic form factors which, in the notation of Bauer, Stech and Wirbel (
henceforth BSW ) [2], are $F_1^{BK}(m^2_{J/\Psi})$ for K production and
$A_1^{BK^{*}}(m^2_{J/\Psi})$, $A_2^{BK^{*}}(m^2_{J/\Psi})$ and
$V^{BK^{*}}(m^2_{J/\Psi})$ for $K^{*}$ production.
We emphasize that, in computing the decay matrix element, we use the
factorization assumption as described in [2].
The colour-suppressed amplitudes are proportional to the phenomenological
dimensionless parameter ${\it \Large a}_{2}$ [2].
The resulting branching ratio for the processes in (2) can be conveniently
written as,
\beqa
BR(B \hs{1mm} \lra \hs{1mm} J/\Psi + K ) & = & N \hs{1mm}|V_{cb}|^2 \hs{1mm}
a_2^2 \hs{1mm} |F_1^{BK}(m^2_{J/\Psi})|^2 \hs{2mm}(10^{12}s^{-1} \tau_B)
\label{eq3} \\
& & \no \\
BR(B \lra J/\Psi + K^{*} )_{\lambda\lambda} & = & N^{*} \hs{1mm} |V_{cb}|^2
\hs{1mm} a_2^2 \hs{1mm} |A_1^{BK^*}(m^2_{J/\Psi})|^2 \hs{1mm}
\Sigma_{\lambda\lambda} \hs{2mm} (10^{12}s^{-1} \tau_B) \label{eq4}
\eeqa
where the polarization-dependent quantity $\Sigma_{\lambda\lambda}$ has the
structure

\beq
\Sigma_{LL} = ( a - bx )^2 \hs{30mm} \Sigma_{TT} = 2 ( 1+ c^2 y^2 ),
\label{eq5}
\eeq
L and T standing for longitudinal and transverse for both final vector mesons.
The other quantities in Eqs.(3)-(5) are defined as follows. The normalization
constants $N$ and $N^{*}$ depend on the masses, on the CKM factor $|V_{cs}|$ =
0.974 and on the leptonic decay constant $f_{J/\Psi}$ estimated from $J/\Psi
\Lra e^{+}e^{-}$ decay to be  $f_{J/\Psi}$ = 382 $MeV$.
The actual values of $N$ and $N^*$ depend on whether the neutral or charged
strange meson is produced, however the difference appears at the level of 1
part in $10^3$ hence we ignore the mass differences between $K^o$ and $K^{+}$,
$K^{*o}$ and $K^{*+}$. We obtain :

\beq
N = 10.845 \hs{2mm}( \frac{f_{J/\Psi}}{382 MeV} )^2 \hs{30mm}
N^{*} = 11.723 \hs{2mm} ( \frac{f_{J/\Psi}}{382 MeV} )^2        \label{eq6}
\eeq

In (5) the dimensionless coefficients $a$, $b$ and $c$ are given by

\beqa
,a & = & \frac{m_B^2 - m^2_{K^*} - m^2_{J/\Psi}}{2 m_{K^*} m_{J/\Psi} } \  \no
\\
b & = & \frac{ 2 K^2 m_B^2}{m_{K^*} m_{J/\Psi} (m_B + m_{K^{*}})^2 } \
\label{eq7} \\
c & = & \frac{2 K m_B}{ ( m_B + m_{K^{*}})^2 }, \ \no
\eeqa
where $K$ is the C$\cdot$M momentum
\beq
2 K m_B = \sqrt{(m_B^2 - m^2_{K^*} - m^2_{J/\Psi})^2 - 4 m^2_{K^*} m^2_{J/\Psi}
}. \label{eq8}
\eeq
The coefficients $a$, $b$ and $c$ depend on the masses somewhat more
sensitively than the coefficients $N$ and $N^*$.
However for the fractional polarization, the difference between neutral and
charged modes is negligible, and we shall use

\beq
a = 3.1475 \hs{20mm} b = 1.2969 \hs{20mm} c = 0.4345 \label{eq9}
\eeq

In Eq(5), the parameters $x$ and $y$ are the ratios of hadronic form factors at
the $J/\Psi$ mass

\beq
x = \frac{A_2^{BK^*}(m^2_{J/\Psi}) }{A_1^{BK^*}(m^2_{J/\Psi}) } \ \hs{30mm}
y = \frac{V^{BK^*}(m^2_{J/\Psi}) }{A_1^{BK^*}(m^2_{J/\Psi}) } \ \label{eq10}
\eeq

Let us now consider the following two experimental data from CLEO II [1]

\hs{20mm} (i) the ratio of vector to pseudoscalar production rates
\beq
R = \frac{ \Gamma(B \lra J/\Psi + K^{*}) }{ \Gamma(B \lra J/\Psi + K )} \ =
1.64 \pm 0.34 \label{eq11}
\eeq

\hs{20mm} (ii) the fraction of longitudinal polarization in $B \Lra J/\Psi +
K^{*} $ decay

\beq
\frac{\Gamma_{L}}{\Gamma} \ = \frac{ \Gamma(B \lra J/\Psi + K^{*})_{LL} }{
\Gamma(B \lra J/\Psi + K^* )} \ = 0.84 \pm 0.06 \pm 0.08 \label{eq12}
\eeq

It is important to note here that ARGUS group found $\Gamma_L/ \Gamma > 0.78 $
at $90\% $ C.L [1, 3].

Assuming factorization, the theoretical expressions for $R$ and $\Gamma_L/
\Gamma $ are
\beq
R_{th} = 1.0809 \hs{2mm}
\frac{|A_1^{BK^*}(m^2_{J/\Psi})|^2}{|F_1^{BK}(m^2_{J/\Psi})|^2} \
\left\{ ( a - bx )^2 + 2 ( 1 + c^2 y^2 ) \right\} \label{eq13}
\eeq
and
\beq
\left[ \frac{\Gamma_L}{\Gamma} \  \right]_{th} = \frac{ (a - bx )^2}{ ( a - bx
)^2 + 2 ( 1 + c^2 y^2 ) } \
\label{eq14}
\eeq

We begin by comparing theory with experiment, purely phenomenologically, with a
view to set limits on the parameters that enter expressions (13) and (14).
It is convenient to discuss the ratio $\Gamma_L/ \Gamma $ first.
In Figure 2 we have plotted the fraction $\Gamma_L/ \Gamma $ in the $x, y $
plane.
The curves of Eq.(14) at fixed $\Gamma_L/ \Gamma $ are hyperbolae centered at
the point $x = \frac{a}{b} \ = 2.43$ and $y = 0$ - the second branch of the
hyperbolae not represented in Figure 2 has been disregarded as unphysical
because it would require too large values of $x \hs{2mm} ( x \simeq 2a/b)$ to
fit experiments.
In Figure 2 we have also shown the one standard deviation lower bound
$\Gamma_L/\Gamma > 0.74$ from CLEO II data. We also observe that there exists a
theoretical upper bound $(\Gamma_L/\Gamma)_{th} \leq a^2/(2 + a^2) $ which on
using (9) yields $(\Gamma_L/\Gamma)_{th} \leq 0.832 $ which is clearly
consistent with experiment.
The bounds on $x$ and $y$ set by $\Gamma_L/\Gamma > 0.74$, shown by the shaded
area in Fig. 2, are :

\beq
x \leq 0.59 \hs{30mm} y \leq 1.98  \label{eq15}
\eeq

We assume the ratios $x$ and $y$ to be positive as is implied by all
theoretical models that we are aware of.
Of course, with only one experimental quantity $\Gamma_L/ \Gamma $ it is not
possible to determine $x$ and $y$ separately - the separation of the two
transverse polarizations is necessary for obtaining $y$ - nevertheless the
allowed domain in the $x$, $y$ plane is small enough to constrain the quantity
$\Sigma_{LL} + \Sigma_{TT}$ associated with the unpolarized rate. From the
allowed domain in the $x$, $y$ plane corresponding to $\Gamma_L/ \Gamma  \geq
0.74 $, we get

\beq
7.7 \hs{2mm} \leq \hs{2mm} ( a - bx )^2 + 2 ( 1 + c^2 y^2 ) \hs{2mm} \leq
\hs{2mm} 13.4 \label{eq16}
\eeq

Next we consider $R_{th}$ of Eq.(13) which we write in the form

\beq
R_{th} = 1.0809 \hs{2mm} \frac{( a - bx )^2 + 2 ( 1 + c^2 y^2 )}{z^2} \
\label{eq17}
\eeq
where we have introduced a parameter $z$ defined by

\beq
z \hs{2mm} = \hs{2mm} \frac{F_1^{BK}(m^2_{J/\Psi}) } {A_1^{BK^*}(m^2_{J/\Psi})
} \ \label{eq18}
\eeq

Using CLEO II result for R, Eq.(11), and the  bounds in (16) we obtain the
following allowed range for $z$

\beq
2.05 \hs{2mm} \leq \hs{2mm} z \hs{2mm} \leq \hs{2mm} 3.34  \label{eq19}
\eeq

Having experimental data on $R$ and $\Gamma_L/\Gamma$ and the associated limits
on $x$, $y$ and $z$ derived by us in (15) and (19), we now investigate the
predictions of various theoretical models. We consider five such models :

\vs{3mm}
\hs{10mm} (i) The original BSW model [2]- called BSWI here-where the value of
the form factors is calculated at $q^2 = 0 $ and
extrapolated  to finite $q^2$ using a monopole form for all the form factors
$F_1, A_1, A_2 $ and $V$.

\vs{3mm}
\hs{9mm} (ii) A modified BSW model- called BSWII here- which takes  the values
of the form factors  calculated at $q^2 = 0$ as in BSWI but uses a monopole
extrapolation for $A_1$ and a dipole extrapolation for $F_1, A_2$ and $V$ [4].

\vs{3mm}
\hs{8mm} (iii) The model of Casalbuoni et al [5] and Deandrea et al [6] where
the normalization at $q^2 =0$ is obtained in  a model that combines heavy quark
symmetry with chiral symmetry for light pseudoscalar degrees of freedom and
also introduces light vector degrees of freedom. We call this the CDDFGN model.
Here all form factors are extrapolated with monopole forms as in Ref.[6].

\vs{3mm}
\hs{9mm} (iv) We  consider a heavy quark type model in which both $b$ and $s$
quarks are treated as heavy( we label this model HSQ for 'heavy strange
quark'). Such a model has been  used by Ali et al [7] in studying $D \ra (K,
K^*) l \nu$, $B \ra K^* \gamma$ and $B \ra (K, K^*) l\ol{l}$.
In our calculations, the four form factors are expressed in terms of an
universal Isgur-Wise function where the extrapolation from  the symmetry point
$q^2 = q^2_{max} $ to $q^2 = m^2_{J/\Psi}$ is made by using an improved form of
the relativistic oscillator model as described in [8].

\vs{3mm}
\hs{9mm} (v) A relativistic constituent quark model due to Jaus and Wyler [9]
which uses light-front formalism to compute the form factors in space-like
momentum transfer region. The form factors are then extrapolated to the
time-like region by a particular two-parameter formula, described in [9],
which, for $q^2 =0$, reproduces the value of the form factors and their first
two derivatives.

\vs{3mm}
In Table I we have  tabulated the values of  $R$, $\Gamma_L/\Gamma$, $x$, $y$
and $z$ obtained in these five models. The two first ones have to be compared
with experiments Eqs.(11) and (12), the three last ones with our constraints
(15) and (19).

\vs{5mm}
\begin{center}
\begin{tabular}{|c||c|c|c|c|c|c|} \hline
 & BSWI  & BSWII & CDDFGN  &  HSQ  & JW  &  \\
\hline\hline
R & 4.25 & 1.61 & 1.46 & 8.97 & 2.44 & 1.64 $\pm$ 0.34 \hs{3mm} $^a$ \\
\hline
$\Gamma_L/\Gamma$ & 0.57 & 0.35 & 0.35 & 0.43 & 0.44 & 0.84 $\pm$ 0.10 \hs{3mm}
 $^b$  \\
\hline \hline
$x$ & 1.01 & 1.41 & 1.05 & 1.34 & 1.17 & $\leq $ 0.59 \hs{3mm}  $^c$ \\
\hline
$y$ & 1.20 & 1.77 & 3.24 & 1.34 & 1.91 & $\leq $ 1.98 \hs{3mm}  $^c$ \\
\hline
$z$ & 1.23 & 1.82 & 2.60 & 0.75 & 1.64 & 2.05 $\leq \hs{2mm} z \leq \hs{2mm} $
3.34 \hs{3mm}  $^d$  \\
\hline
\end{tabular} \\

{\small $a$: Eq.(11), $b$: Eq.(12), $c$: Eq.(15), $d$: Eq.(19) }

\vs{5mm}
$\ul{\rm Table \hs{2mm} 1.}$
\end{center}

Both BSWI and HSQ models are clearly ruled out by the ratio of the rates $R$,
the latter faring far worse than the former$^{\dagger}$
\footnote{
$\dagger$ \hs{2mm} This is not surprising as such a  model is also known [7] to
overestimate the semileptonic rates $D \ra (K, K^*) l\nu $ by a factor of 2.
Our analysis clearly demonstrates that treating the $s$ quark as heavy is a
poor approximation. }.
JW model overestimates R moderately.
The remaining two models BSWII and CDDFGN are consistent with data on $R$
within experimental errors.

However the situation takes a dramatic turn for $\Gamma_L/\Gamma$ where
$\ul{none}$ of the five models considered reproduce the large longitudinal
polarization fraction observed experimentally - equivalently the allowed domain
in the $x$, $y$ plane excludes the values of $x$ and $y$ produced by all of
these models. See Figure 2 where the $x$, $y$ parameters produced by these
models are shown.

We conclude  this part of our discussion by noting that these models,
generally used in the determination of the parameters $a_2$ from $B \ra J/\Psi
+ K $ and  $B \ra J/\Psi + K^* $ data, are unable to explain the large
longitudinal polarization fraction
$\Gamma_L/\Gamma$ observed experimentally.
This might be due to  either a failure of the factorization procedure or to a
wrong choice of hadronic form factors or both.
In either case the value of $|a_2|$ quoted in the literature becomes suspect
and if correct, it might be due to a fortunate accident.

In what follows we still assume factorization even though we are aware of the
fact that the extrapolation from $q^2 =0$ to $q^2 = m^2_{J/\Psi}$ is a large
one and the previously discussed difficulties may have their origin in large
corrections to factorization and/or to $q^2$ extrapolation.

The determination of $|a_2|$ from experiments depends on the lifetime
$\tau_{B}$, the CKM factor $|V_{cb}|$, the leptonic constant $f_{J/\Psi}$ and
the hadronic form factors $F_1, A_1, A_2$ and $V$. Using for $\tau_B$ and
$|V_{cb}|$ a standard set proposed in Ref.[1].
\beq
\tau_B = 1.44 \hs{2mm} 10^{-12} s \hs{30mm} |V_{cb}| = 0.041 \label{eq21}
\eeq
we obtain the following theoretical branching ratios
\beqa
BR(B \lra J/\Psi + K) & = & 2.62 \hs{1mm} \cdot 10^{-2} \hs{2mm} a^2_2 \hs{2mm}
|F_1^{BK}(m^2_{J/\Psi})|^2 \label{eq22} \\
& &  \no \\
BR(B \lra J/\Psi + K^*) & = & 2.84 \hs{1mm} \cdot 10^{-2} \hs{2mm} a^2_2
\hs{2mm} |A_1^{BK^*}(m^2_{J/\Psi})|^2 [ \Sigma_{LL} + \Sigma_{TT} ]
\label{eq23}
\eeqa
The World average of the experimental data are [1]
\beqa
BR(B^o \ra J/\Psi + K^o)  & = &  (0.08 \pm 0.02) \cdot 10^{-2} \no \\
\cr
BR(B^{+} \ra J/\Psi + K^{+}) & = & (0.10 \pm 0.01) \cdot 10^{-2} \no \\
\cr
BR(B^o \ra J/\Psi + {K^o}^*) & = & (0.15 \pm 0.08) \cdot 10^{-2}  \label{eq24}
\\
\cr
BR(B^{+} \ra J/\Psi + K^{*+})  & = & (0.17 \pm 0.05) \cdot 10^{-2} \no
\eeqa
{}From the intersection of one standard deviation domain we obtain
\beqa
|a_2| F_1^{BK}(m^2_{J/\Psi}) & = & 0.190 \pm 0.005 \label{eq25} \\
\cr
|a_2| A_1^{B{K^{*}}}(m^2_{J/\Psi}) [ \Sigma_{LL} + \Sigma_{TT} ]^{1/2} & = &
0.229 \pm 0.023 \label{eq26}
\eeqa
Using our limits (16) set by the polarization measurement, we obtain
\beq
0.056 \leq |a_2| A_1^{B{K^{*}}}(m^2_{J/\Psi}) \leq 0.091 \label{eq27}
\eeq
The ranges (19) on the parameter $z$ can be recovered by eliminating $|a_2|$
between (24) and (26).

The five models we have discussed here being unable to account for both  CLEO
II data on $R$ and $\Gamma_L/\Gamma$, it is interesting to seek some other way
to obtain $B \ra K$ and $B \ra K^*$ hadronic form factors with as little model
dependence as possible.
This can be achieved by relating $B \ra (K, K^*)$ form factors to the measured
$D \ra (K, K^*)$ form factors in, at least, two different ways: one, using the
method of heavy quark symmetry combined with chiral symmetry incorporating  the
light vector degrees of freedom [5, 6, 10], and, second, by using the method of
Isgur and Wise [11].
We choose the latter as it is based on general principles of scaling and heavy
flavour symmetry, though it relates the form factors at the same heavy flavour
velocity - rather than the same momentum -  transfer.
The chiral method on the other hand depends on what is put in the theory; as an
example [10], whether an axial heavy meson is built-in or not makes an
important difference to the form factors, particularly $A_2$.
In addition, one has to assume flavour independence of the strong couplings
involving $B$ and $D$ mesons, these couplings being relevant to the computation
of the pole contributions to the form factors [10].

The flavour symmetry between $b$ and $c$ quarks allows us to derive relations
between the
$B \ra (K, K^*)$ and $D \ra (K, K^*)$ hadronic form factors at the same
velocity transfer, though at different momentum transfers.
Calling $t_B (t_D)$ the value of the squared momentum transfer, $q^2$, for $B (
D )$ form factors, we obtain the following kinematic relations
\beqa
v_b \cdot v_K & = & v_c \cdot v_K \hs{10mm} {\rm or} \hs{10mm} \frac{t_B -
m^2_K - m^2_b}{m_b} \ = \frac{t_D - m^2_K - m^2_c}{m_c} \ \label{eq28} \\
\cr
v_b \cdot v_{K^*} & = & v_c \cdot v_{K^*} \hs{10mm} {\rm or} \hs{10mm}
\frac{t_B - m^2_{K^*} - m^2_b}{m_b} \ = \frac{t_D - m^2_{K^*} - m^2_c}{m_c} \
\label{eq29}
\eeqa
Using Isgur-Wise method [11], we get
\beqa
F_1^{BK}(t_B) & = &  [ \frac{\alpha_s(m_b)}{\alpha_s(m_c)} ]^{-6/25} \left\{
\frac{m_b + m_c}{2 \sqrt{m_b m_c} } \ F_1^{DK}(t_D) - \frac{m_b - m_c}{2
\sqrt{m_b m_c}} \ \frac{m_D^2 -m_K^2}{t_D} \ [F_0^{DK}(t_D) - F_1^{DK}(t_D)]
\right\} \no \\
\cr
A_1^{B{K^*}}(t_B) & = &  [ \frac{\alpha_s(m_b)}{\alpha_s(m_c)} ]^{-6/25}
\frac{m_D + m_{K^*}}{m_B + m_{K^*}} \ \sqrt{ \frac{m_b}{m_c} \ }
A_1^{D{K^*}}(t_D) \label{eq30} \\
\cr
A_2^{B{K^*}}(t_B) & = &  [ \frac{\alpha_s(m_b)}{\alpha_s(m_c)} ]^{-6/25}
\frac{1}{2} \  \sqrt{ \frac{m_c}{m_b} \ }
\left\{ ( 1 + \frac{m_c}{m_b} \ ) \frac{m_B + m_{K^*}}{m_D + m_{K^*}} \
A_2^{D{K^*}}(t_D) +
( 1 - \frac{m_c}{m_b} \ ) \frac{m_B + m_{K^*}}{t_D} \  \right.
\cr
& & \hs{15mm} \left.  \cdot [ 2 m_{K^*} A_0^{D{K^*}}(t_D) - ( m_D + m_{K^*} )
A_1^{D{K^*}}(t_D) + ( m_D - m_{K^*} ) A_2^{D{K^*}}(t_D)] \right\}   \no \\
\cr
V^{B{K^*}}(t_B) & = &  [ \frac{\alpha_s(m_b)}{\alpha_s(m_c)} ]^{-6/25}
\frac{m_B + m_{K^*}}{m_D + m_{K^*}} \ \sqrt{ \frac{m_c}{m_b} \ }
V^{D{K^*}}(t_D) \no
\eeqa
We remark that the right hand sides of $F_1^{BK}$ and $A_2^{B{K^*}}$ are not
singular at $t_D = 0$ because of the normalization constraints [2]

\beqa
& & F_0^{DK}(0) - F_1^{DK}(0)  =  0  \no \\
& &  \label{eq31} \\
& & 2 m_{K^*} A_0^{D{K^*}}(0) - ( m_D + m_{K^*} ) A_1^{D{K^*}}(0) + ( m_D -
m_{K^*} ) A_2^{D{K^*}}(0)  =  0 \no
\eeqa
We are interested in $B \ra (K, K^*)$ form factors at the $J/\Psi$ mass, $t_B =
m^2_{J/\Psi}$, while the input is the experimental values of the $D \ra (K,
K^*)$ form factors at $t_D = 0$.
Two sets of $(t_B, t_D)$ of interest are $(m^2_{J/\Psi}, t^o_D)$ and $(t_B^o,
0)$, where $t_D^o$ and  $t_B^o$ are computed from either (27) or (28) depending
on whether one is considering $B \ra K$ or $B \ra K^*$ transition.
If we use the first set, we need to extrapolate $D \ra (K, K^*)$ form factors
from $t_D^o$ to 0 where
experimental information exists.
With the second set we need to extrapolate $B \ra (K, K^*)$ form factors from
$t_B^o$ to $m^2_{J/\Psi}$.
The resulting $B \ra (K, K^*)$ form factors at $J/\Psi$ mass must be the same
if the two extrapolations were to be consistent.

We have assumed for all form factors a monopole dependence with the mass poles
as given in Table 2 due to BSW[2].

\vs{4mm}
\begin{center}
\begin{tabular}{|c||c|c|c|c|} \hline
$J^P$ & $0^{+}$   &  $1^{-}$ & $1^{+}$  & $0^{-}$  \\
\hline\hline
$c \ra s$ & 2.60 & 2.11 & 2.53 & 1.97  \\
\hline
$b \ra s$ & 5.89 & 5.43 & 5.82 & 5.38  \\
\hline
\end{tabular}

\vs{5mm}
$\ul{\rm Table \hs{2mm} 2.} $ \\
( Pole masses in GeV )
\end{center}

The compatibility of the two extrapolation procedures for $D \ra (K, K^*)$ and
$B \ra (K, K^*)$ form factors is very satisfactory. The two would be identical
if the pole masses $\Lambda_B$ in the $B$ sector and $\Lambda_D$ in the $D$
sector for a given $J^P$ were related by
\beq
\Lambda_B^2 = t_B^o + \frac{m_b}{m_c} \ \Lambda_D^2 \label{eq35}
\eeq
Such relations are satisfied to a good accuracy by the poles of Table 2.

Using the experimental input on $D \ra (K, K^*)$ form factors from [12]

\beqa
F^{DK}_1(0) & = & 0.77 \pm 0.04  \hs{20mm} A^{DK^*}_1(0) = 0.61 \pm 0.05 \no \\
& &  \label{eq32} \\
A^{DK^*}_2(0) & = & 0.45 \pm 0.09  \hs{20mm} V^{DK^*}(0) = 1.12 \pm 0.16 \no
\eeqa
we obtain, using Eq.(29) with $m_b = 5.0 \hs{2mm} GeV$, $m_c = 1.5 \hs{2mm}
GeV$,
\beqa
F^{BK}_1(m^2_{J/\Psi}) & = & 0.75 \pm 0.05  \hs{20mm} A^{BK^*}_1(m^2_{J/\Psi})
= 0.43 \pm 0.04 \no \\
& &  \label{eq33} \\
A^{BK^*}_2(m^2_{J/\Psi}) & = & 0.42 \pm 0.07 \hs{20mm} V^{BK^*}(m^2_{J/\Psi}) =
1.10 \pm 0.17 \no
\eeqa
The results quotes here are the union of the two ranges using extrapolations in
$t_D$ and $t_B$ with symmetrical errors.

The value of $|a_2|$ is most accurately obtained from $B \ra J/\Psi + K $ data
and using equations (24) and (33) we get
\beq
|a_2| = 0.252 ^{+0.025}_{-0.022} \label{eq34}
\eeq
We observe that the result(34) is very similar to the one obtained with the
models BSWII anf CDDFGN which reproduce correctly the ratio $R$.
\beqa
|a_2| & = & 0.228 \pm 0.006  \hs{35mm} {\rm BSWII} \no \\
& &  \label{eq36} \\
|a_2| & = & 0.262 \pm 0.007  \hs{34mm} {\rm CDDFGN} \no
\eeqa
However the form factors of (33) also lead us to
\beq
x = 0.97 \pm 0.26  \hs{30mm} y = 2.56 \pm 0.64  \label{eq37}
\eeq
The corresponding point is shown on Figure 2 and is clearly outside the allowed
domain defined by $\Gamma_L/\Gamma > 0.74 $.
 From the range $x$ and $y$ in (36),
the value of the longitudinal polarization fraction is found to be :
\beq
\frac{\Gamma_L}{\Gamma} = 0.45^{+0.13}_{-0.17} \label{eq38}
\eeq
which differs from the experimental value (12) by more than two standard
deviations.
We observe that BSWII and CDDFCN models both predict a value of
$\Gamma_L/\Gamma = 0.35 $
which is five standard deviation off the experimental value.
Thus the Isgur-Wise procedure to derive the $B \ra (K, K^*)$ form factors from
the $D \ra (K, K^*)$ ones
also fails in producing form factors consistent with  $(\Gamma_L/\Gamma)$ data.

The problem may be due to the procedure, and we shall discuss this point later,
or it may well be with the input which is the experimental determination of $D
\ra (K, K^*)$ form factors, in particular $A^{DK^*}_2$ and $V^{DK^*}$.
The history of $A^{DK^*}_2$ is chequered to say the least.
The earlier E-691 determination [13] of $A^{DK^*}_2(0)$ was consistent with
zero and implied a large longitudinal polarization in semileptonic $D \ra K^*l
\nu_l $ decays. Later measurements by E-653[14] and E-687[15] collaborations
led to a larger
 $A^{DK^*}_2(0)$ and consequently smaller values of the longitudinal
polarization.
It is these larger values of  $A^{DK^*}_2(0)$ which are being reflected to
larger values of  $A^{BK^*}_2(m^2_{J/\Psi})$
through the Isgur-Wise procedure.
The value of $V^{BK^*}(m^2_{J/\Psi})$ is also too high, however the error in
the input value $V^{DK^*}(0)$
is also large and with a generous treatment of errors, $V^{BK^*}(m^2_{J/\Psi})$
could well be within the $y$ bound (15) set by us.
The major problem is with $A^{BK^*}_2(m^2_{J/\Psi})$ or equivalently the
parameter $x$.

Returning now to a discussion of theoretical procedure in deriving $B \ra (K,
K^*)$ form factors from $D \ra (K, K^*)$ ones, we wish to emphasize two sources
of uncertainties:

\hs{10mm} (i) the choice of $b$ and $c$ quark masses, and

\hs{10mm} (ii) the type of $q^2$-extrapolation employed.

In order to obtain an estimate of the theoretical error due to the choice of
quark masses, we repeat our calculation
replacing $m_b$ by $m_B$ and $m_c$ by $m_D$.
The resulting form factors are,
\beqa
F^{BK}_1(m^2_{J/\Psi}) & = & 0.66 \pm 0.04  \hs{20mm} A^{BK^*}_1(m^2_{J/\Psi})
= 0.37 \pm 0.03 \no \\
& &  \label{eq39} \\
A^{BK^*}_2(m^2_{J/\Psi}) & = & 0.43 \pm 0.08 \hs{20mm} V^{BK^*}(m^2_{J/\Psi}) =
1.09 \pm 0.16 \no
\eeqa

Comparing (33) with (38) we note that only $F^{BK}_1$ and $A^{BK^*}_1$ are
affected to any extent by the choice of quark masses; however, within errors,
the results are consistent.

The type of $q^2$-dependence of the hadronic form factors poses a more serious
problem.
The $D \ra (K, K^*)$ experiments have been analysed [12] with monopole form
factors, for that reason we have used a monopole form  for all $D \ra (K, K^*)$
and $B \ra (K, K^*)$ form factors.
This ansatz, that all form factors can be extrapolated using monopole forms,
could be incorrect at least for some form factors.
We already know that for heavy to heavy quark transition, consistency with the
heavy quark limit requires the $q^2$-dependence of $F_1(q^2)$, $V(q^2)$ and
$A_2(q^2)$ to differ from that of $A_1(q^2)$ by an additional pole factor [4].
However, for heavy to light quark transitions, to the best of our knowledge,
there is no convincing argument as to the type of $q^2$-dependence to be
employed.
At the other extreme, $QCD$ sum rules in the $D \ra K^{*}$ transition seem to
indicate that for the hadronic form factor $A_1(q^2)$, a single monopole form
may be questionable, and instead of growing with positive $q^2$, $A_1(q^2)$ may
decrease [16].
Such a feature may change drastically the analysis and may help in
understanding our phenomenological results (15) and (19).

\vs{5mm}
In summary,
the polarization measurement in $B \ra J/\Psi K^* $ by CLEO II and ARGUS groups
appears as decisive input in the analysis of this class of decays.
It sets very strigent limits on our ratios $x$ and $y$ which up to now are not
theoretically understood.
The six models we have selected as representative ones are unable to reproduce
the data by many standard deviations.
It is clear that the confirmation of the polarization measurement in $B \ra
J/\Psi K^*$ is urgently needed.
On the other hand, a more refined study of the hadronic form factors in $B \ra
K( K^* ) $ transitions is obviously necessary.
It would be interesting to have a reliable estimate of the corrections to the
factorization assumption which might be part of the actual difficulty to
explain experiment.
In any case, we must be very cautious about the determination of the parameter
$a_2$ obtained from these decay modes.

\vspace{0.5cm}
{\bf ACKNOWLEDGMENTS}
\vs{5mm}

A. N. K would like to thank the Laboratoire de Physique Th\'eorique et Hautes
Energies at Paris for their hospitality and the Natural Sciences and
Engineering Research Council of Canada for a research grant which partly
supported this research.
We thank Mr. Yong-Yeon Keum for checking our calculations.

\newpage

\vs{20mm} \large
{\bf References} \normalsize

\vs{5mm}

{\bf [1]} T. E. Browder, K. Honscheid and S. Playfer, Report CLNS 93/1261,
1994.

\hs{5mm} To appear in $\ul{\rm B \hs{2mm} Decays}$, 2nd. Edition, Ed. S. Stone,
World Scientific, Singapore.

{\bf [2]} M. Bauer, B.Stech and M.Wirbel, Z. Phys. $\ul{\rm C 34}$, 103(1987);

\hs{5mm} M. Wirbel, B.Stech and M. Bauer, Z. Phys. $\ul{\rm C 29}$, 637(1985).

{\bf [3]} P. Krizan, in Proceedings of the Rencontres de Moriond 1992, Ed. J.
Tran Thanh

\hs{5mm} Van Editions Frontieres; H. Schroder, in Proceedings of the 25th
International

\hs{5mm} Conference on High Energy  Physice, Eds, K. K. Phua and Y. Yamaguchi,
Vol. II,

\hs{5mm} P846, World Scientific, Singapore, 1991.

{\bf [4]} See, for example, M. Neubert, V. Rieckert, B. Stech and Q. P. Xu in
$\ul{\rm Heavy \hs{2mm} Flavours}$,

\hs{5mm} Eds. A. J. Buras and M. Lindner, World Scientific, Singapore, 1992.

{\bf [5]} R. Casalbuoni, A. Deandrea, N. Di Bartolomeo, R. Gatto, F. Feruglio
and

\hs{5mm} G. Nardulli, Phys. Lett. $\ul{\rm B 292}$
,371(1992)

{\bf [6]} A. Deandrea, N. Di Bartolomeo, R. Gatto and G. Nardulli, Phys. Lett.
$\ul{\rm B 318}$,

\hs{5mm} 549(1993)

{\bf [7]}A. Ali and T. Mannel, Phys. Lett. $\ul{\rm B 264}$, 447(1991);
$\ul{\rm B 274}$, 256(1992)

\hs{5mm} A. Ali, T. Ohland and T. Mannel, Phys. Lett. $\ul{\rm B 298}$,
195(1993)

{\bf [8]} M. Neubert and V. Rieckert, Nucl. Phys. $\ul{\rm B 285}$, 97(1992)

{\bf [9]} W. Jaus, Phys. Rev. D$\ul{41}$, 3394(1990);

\hs{5mm} W. Jaus and D. Wyler, Phys. Rev. D$\ul{41}$, 3405(1990).

{\bf [10]} J. Schechter and A. Subbaraman, Phys. Rev. D$\ul{48}$, 332(1993);

\hs{5mm} A. N. Kamal and Q. P. Xu,  Phys. Rev. D$\ul{49}$, 1526(1994).

{\bf [11]} N. Isgur and M. B. Wise, Phys. Rev. D$\ul{42}$, 2388(1990).

{\bf [12]} M. S. Witherell, Invited talk at International Symposium on Lepton
and Photon

\hs{5mm} Interactions at High Energies, Cornell, Ithaca, N. Y.(1993)

\hs{5mm} Report UCSB-HEP-93-06 (to be Published).

{\bf [13]} J. C. Anjos et al, E-691 Collaboration, Phys. Rev. Lett. $\ul{62}$,
722(1989);

\hs{5mm} $\ul{65}$, 2630(1990).

{\bf [14]} K. Kodama et al, E-653 Collaboration, Phys. Lett. $\ul{\rm B 274}$,
246(1992)

{\bf [15]} P. L. Frabetti  et al, E-687 Collaboration, Phys. Lett. $\ul{\rm B
307}$, 262(1993)

{\bf [16]} S. Narison, Private communication.


\newpage
\large
{\bf Figure captions}\\

\normalsize
\vspace{0.5cm}

{\bf Figure 1.} The colour-suppressed  Feynman diagram. \\

{\bf Figure 2.} The polarization fraction $\Gamma_L/\Gamma$ in the $x, y $
plane. The solid curves correspond  to fixed values of $\Gamma_L/\Gamma$ as
indicated.
The dashed curve is the theoretical upper bound $\Gamma_L/\Gamma = 0.832$.
The shaded region is the allowed region corresponding to $\Gamma_L/\Gamma >
0.74$.
The six points correspond to the predictions of the models defined in the text,
A: BSWI, B: BSWII, C: CDDFGN, D: HSQ, E: JW, F: Isgur-Wise procedure Eq.(36).
Errors are shown when available.

\end{document}